\documentclass{elsart}

 \usepackage{epsfig}

\usepackage{amssymb}
\journal{Computer Physics Communications}

\newcommand{\jpb}{J. Phys. B }
\newcommand{\PR}{Phys. Rev. }
\newcommand{\PRL}{Phys. Rev. Lett. }

\newcounter{bla}

\begin{document}
\begin{frontmatter}

\title{{\sc ClassSTRONG}: Classical simulations of Strong Field processes}
\author[auburn,icfo]{M. F. Ciappina\corauthref{cor}} and
\corauth[cor]{Corresponding author.}
\ead{mfc0007@auburn.edu}
\author[clpu]{J. A. P\'erez-Hern\'andez}
\author[icfo,icrea]{M. Lewenstein}


\address[auburn]{Department of Physics\\
 Auburn University, Auburn, Alabama 36849, USA}
\address[icfo]{ICFO-Institut de Ci\`ences Fot\`oniques, \\
08860 Castelldefels (Barcelona), Spain}
\address[clpu]{Centro de L\'aseres Pulsados (CLPU), Parque Cient\'{\i}fico, 37185 Villamayor, Salamanca, Spain}
\address[icrea]{ICREA-Instituci\'o Catalana de Recerca i Estudis Avan\c{c}ats, Lluis
Companys 23, 08010 Barcelona, Spain}

\begin{abstract}
A set of Mathematica functions is presented to model classically two of the most important processes in strong field physics, namely high-order harmonic generation (HHG) and above-threshold ionization (ATI). Our approach is based on the numerical solution of the Newton-Lorentz equation of an electron moving on an electric field and takes advantage of the symbolic languages features and graphical power of Mathematica. Similarly as in the Strong Field Approximation (SFA), the effects of atomic potential on the motion of electron in the laser field are neglected. The SFA has proven to be an essential tool in strong field physics in the sense that it is able to predict with great precision the harmonic (in the HHG) and energy (in the ATI) limits. We have extended substantially the conventional classical simulations, where the electric field is only dependent on time, including spatial nonhomogeneous fields and spatial and temporal synthesized fields. Spatial nonhomogeneous fields appear when metal nanosystems interact with strong and short laser pulses and temporal synthesized fields are routinely generated in attosecond laboratories around the world. Temporal and spatial synthesized fields have received special attention nowadays because they would allow to exceed considerably the conventional harmonic and electron energy frontiers. Classical simulations are an invaluable tool to explore exhaustively the parameters domain at a cheap computational cost, before massive quantum mechanical calculations, absolutely indispensable for detailed analysis, are performed. 

\begin{flushleft}
PACS: 34.10.+x; 34.50.-s; 34.50.Fa; 34.90.+q
\end{flushleft}

\begin{keyword}
Strong Field Physics \sep High-order harmonics generation \sep Above threshold ionization \sep classical simulations
\end{keyword}

\end{abstract}

\end{frontmatter}

\newpage
{\bf PROGRAM SUMMARY}

\begin{small}
\noindent
{\em Manuscript Title:} {\sc ClassSTRONG}: Classical simulations of Strong Field processes                 \\
{\em Authors:} M. F. Ciappina, J. A. P\'erez-Hern\'andez and M. Lewenstein                                                 \\
{\em Program Title:} ClassSTRONG                                          \\
{\em Journal Reference:}                                      \\
{\em Catalogue identifier:}                                   \\
{\em Licensing provisions:} none                                   \\
{\em Programming language:} Mathematica                   \\
{\em Computer:} Single machines using Linux or Windows (with cores with any clock speed, cache memory and bits in a word)                                              \\
{\em Operating system:} Any OS that supports Mathematica. The notebooks have been tested under Windows and Linux and with versions
6.x, 7.x, 8.x and 9.x                                    \\
{\em Number of processors used:} any number supported by Mathematica                             \\
{\em Supplementary material:} Notebooks with different examples are provided                                 \\
{\em Keywords:} Strong Field Physics, High-order harmonics generation, Above threshold ionization, classical simulations, Mathematica  \\
{\em PACS:} 34.10.+x; 34.50.-s; 34.50.Fa; 34.90.+q
\\
{\em Classification:}                                          \\
{\em External routines/libraries: } RootSearch.m                                      \\
{\em Subprograms used:}                                       \\

{\em Nature of problem:} The Mathematica functions model high-order harmonic generation (HHG) and above-threshold ionization (ATI) using the classical equations of motion of an electron moving in an oscillating electric field. In Strong Field Physics HHG and ATI represent two of the most prominent examples of the non-perturbative interaction between strong laser sources and matter. In HHG an atomic or molecular bound electron is put into the continuum by the external laser electric field. Due to the oscillatory nature of the electromagnetic radiation, the electron is steered back and recombines with the parent ion converting its kinetic energy as high energy photons. For ATI the electron is laser-ionized in the same way as in HHG, but in its return it is elastically rescattered by the parent ion gaining even more kinetic energy. We incorporate functions for different laser pulse envelopes, namely sine-squared, gaussian, trapezoidal and numerically defined by the user. In addition we relax the assumption of spatial homogeneity of the laser electric field, allowing weak spatial variations with different functional forms. Finally we combine spatial and temporal synthesized laser fields to produce HHG and ATI. For all the cases the functions allow the extraction of pre-formatted graphs as well as raw data which can be used to generate plots with other graphical programs.  \\

  
{\em Solution method:}  The functions employ the numerical solution of the Newton-Lorentz equation for an electron moving in a spatial and temporal varying electric field to calculate the energy and harmonic spectra features of HHG and ATI.  Our approach neglects any magnetic effect. \\
   \\
{\em Restrictions:} No restrictions.\\
   \\
{\em Unusual features:} none\\
   \\
\newpage
{\em Additional comments:}  The set consists of the following 16 notebooks:
\begin{itemize}
\item{{\tt HHGSin2.nb} - 
This notebook includes functions to calculate the high-order harmonic spectra features, both in terms of harmonic order and energy in eV, for atoms interacting with laser pulses with sine-squared envelopes.}
\item{{\tt HHGGauss.nb} - 
This notebook includes functions to calculate the high-order harmonic spectra features, both in terms of harmonic order and energy in eV, for atoms interacting with laser pulses with gaussian envelopes.}
\item{{\tt HHGTrap.nb} - 
This notebook includes functions to calculate the high-order harmonic spectra features, both in terms of harmonic order and energy in eV, for atoms interacting with laser pulses with trapezoidal envelopes. }
\item{{\tt HHGUser.nb} - 
This notebook includes functions to calculate the high-order harmonic spectra features, both in terms of harmonic order and energy in eV, for atoms interacting with laser pulses defined by the user.}
\item{{\tt ATISin2.nb} - 
This notebook includes functions to calculate the above-threshold ionization spectra features, both in terms of electron energy in eV and $U_p$ units, for atoms interacting with laser pulses with sine-squared envelopes.}
\item{{\tt ATIGauss.nb} - 
This notebook includes functions to calculate the above-threshold ionization spectra features, both in terms of electron energy in eV and $U_p$ units, for atoms interacting with laser pulses with gaussian envelopes.}
\item{{\tt ATITrap.nb} - 
This notebook includes functions to calculate the above-threshold ionization spectra features, both in terms of electron energy in eV and $U_p$ units, for atoms interacting with laser pulses with trapezoidal envelopes.}
\item{{\tt ATIUser.nb} - 
This notebook includes functions to calculate the high-order harmonic spectra features, both in terms of harmonic order and energy in eV, for atoms interacting with laser pulses defined by the user.}
\item{{\tt HHGTemporal.nb} - 
This notebook includes functions to calculate the high-order harmonic spectra features, both in terms of harmonic order and energy in eV, for atoms interacting with temporal synthesized laser pulses.}
\item{{\tt ATITemporal.nb} - 
This notebook includes functions to calculate the above-threshold ionization spectra features, both in terms of electron energy in eV and $U_p$ units, for atoms interacting with temporal synthesized laser pulses.}
\item{{\tt HHGLinear.nb} - 
This notebook includes functions to calculate the high-order harmonic spectra features, both in terms of harmonic order and energy in eV, for atoms interacting with spatial inhomogeneous laser pulses (linear case).}
\item{{\tt ATILinear.nb} - 
This notebook includes functions to calculate the above-threshold ionization spectra features, both in terms of electron energy in eV and $U_p$ units, for atoms interacting with spatial inhomogeneous laser pulses (linear case).}
\item{{\tt HHGExp.nb} - 
This notebook includes functions to calculate the high-order harmonic spectra features, both in terms of harmonic order and energy in eV, for atoms interacting with spatial inhomogeneous laser pulses (exponential case).}
\item{{\tt ATIExp.nb} - 
This notebook includes functions to calculate the above-threshold ionization spectra features, both in terms of electron energy in eV and $U_p$ units, for atoms interacting with spatial inhomogeneous laser pulses (exponential case).}
\item{{\tt HHGTemporal\&Spatial.nb} - 
This notebook includes functions to calculate the high-order harmonic spectra features, both in terms of harmonic order and energy in eV, for atoms interacting with temporal and spatial synthesized laser pulses.}
\item{{\tt ATITemporal\&Spatial.nb} - 
This notebook includes functions to calculate the above-threshold ionization spectra features, both in terms of electron energy in eV and $U_p$ units, for atoms interacting with temporal and spatial synthesized laser pulses.}
\end{itemize}

All the notebooks use the Mathematica package {\tt RootSearch.m} developed by Ted Ersek (see e.g.~\cite{rootsearch} for more details).

{\em Running time:} Computational times vary according to the number of points required for the numerical solution of the Newton-Lorentz equation and of the complexity of the spatial and temporal driving laser electric field. The typical running time is several minutes, but it can be larger for large number of optical cycles and spatially and temporal complex laser electric fields.\\
   \\
\end{small}

\newpage


\hspace{1pc}

\section{Introduction}

By exposing an atom or molecule to an intense and short laser pulse it is possible, for instance, to release one of the outermost electrons to the continuum. This laser-ionized electron will move subsequently in a combined field formed by the time oscillating laser electric field and the residual atomic or molecular Coulomb potential. The theoretical treatment of this non-perturbative laser-matter interaction represents a formidable task from a computational viewpoint, even for the simplest systems. There has been, however, important advances in the development of algorithms and the increasing availability of computational power have made it possible to solve the time-dependent Schr\"odinger equation (TDSE) numerically, but only in a restrictive range of field strengths and frequencies. Consequently, approximate theories and alternative approaches still play an important role in both the solution of the problems under study and in the understanding of the main features of the laser-induced phenomena.

One of the most important concepts and that it has driven the attosecond science was the discovery of the so-called 
coherent electron-ion collisions induced by a strong laser field, most commonly referred to as \textit{re-collisions}~\cite{corkum}. This idea
evolved starting from the numerical experiments of Schafer et al.~\cite{schafer}. The work of Corkum~\cite{corkum}, indeed, built on earlier contributions~\cite{brunel,kuchiev,corkum1989} and it has in turn been widely used to classically model different laser driven phenomena.

We can summarize the classical picture of strong-field-induced ionization electron dynamics as follows. Once an atomic or molecular electron is freed to the continuum by the laser, it finds itself moving in the strong oscillating laser field. This configures the so-called strong field approximation, where the main assumption is that any influence of the atomic or molecular potential is neglected. Using the classical equations of motion it is possible to show that, within one or few cycles after ionization, the oscillating electron can be driven back by the laser field to re-encounter the parent ion. At the time this re-encounter, commonly referred to as re-collision, takes place, the electron can initiate several phenomena:  scatter
elastically (diffract), scatter inelastically (excitation or multiple ionization of the parent ion or molecule),
or radiatively recombine into one of the residual atomic or molecular ion's empty states (if particular conditions are met, coherent harmonic radiation can be generated). The classical approach is usually known as the three-step
model, or the simple man's model~\cite{corkum}. 

If the electron elastically rescatters, gains energy and continues its travel to the detector we talk about above-threshold ionization (ATI) (see~\cite{miloreview}). The ATI phenomenon receives special attention due to the fact that the ATI yield is highly sensitive to the so-called carrier envelope phase, and consequently it constitutes a very robust pulse characterization tool. On the other hand if the laser-ionized electron radiatively recombines with the exact same state that the electron has left from, then the phase of the emitted radiation is the same from one atom to another of the sample, leading to the generation of coherent radiation of multiple frequencies. This process is known as high harmonic generation (HHG) and it is the basis for the production of attosecond pulses amongst other uses (see e.g.~\cite{lein}).

The rest of this contribution is organized as follows. In the next section we focus on the definitions of the functions used for the classical simulations of high-order harmonic generation (HHG) driven by an ample of set of laser sources, ranging from the conventional (spatially homogeneous) to more complex (spatially nonhomogeneous of different kinds, temporal synthesized and spatial and temporal synthesized) ones, including different pulses envelopes. Next we treat above threshold ionization (ATI) in a similar way. Finally we conclude with a short summary.
 
\section{High-order harmonic generation (HHG)}
\subsection{High-order harmonic generation (HHG) driven by spatial homogeneous fields}
We start by modeling high-order harmonic generation (HHG) in atoms and molecules driven by sine-squared laser pulses. The Mathematica
functions corresponding to this case will be described in detail and for the rest of the presented cases we only point out the differences and similarities.

We need a couple of initialization commands before to start the definition of the Mathematica functions, namely, \\
{\tt  SetDirectory[NotebookDirectory[]]} \\
and \\
{\tt <<RootSearch.m}\\
The first command sets the working directory to the {\tt  NotebookDirectory[]}, i.e. where the current notebook is located. The second initialization line loads the package {\tt  RootSearch.m} using the {\tt  Get} ($<<$) command. Please note that the {\tt  RootSearch.m} package should to be located in the same directory where the notebooks are. For details about the {\tt  RootSearch.m} see e.g.~\cite{rootsearch}.\\

The first function we define will be used to generate both a plot of the laser electric field and raw data that could be employed by external programs for postprocessing. The function is named {\tt PulseSin2} and the variables needed for its calling are:
\begin{itemize}
\item{The laser intensity, $I$, in W/cm$^{2}$}
\item{The laser wavelength, $\lambda$, in nm}
\item{The carrier envelope phase, $\phi$}
\item{The total number of laser cycles $n_p$}
\item{The number of points $n_{points}$ when the option {\tt Data} is used (optional)}
\item{Select the option to generate a Mathematica plot ({\tt Picture}), to export a pdf version of the laser electric field plot ({\tt File}) or to save in a file a table with the numerical values of the laser electric field vs. time ({\tt Data})}
\end{itemize}
An example of a call of this function is:

{\tt PulseSin2[1$\times$10$^{{\tt 14}}$, 1800, $\pi$, 10, Option -> "Picture"]}

It produces a plot in the screen of the laser electric field, with the corresponding units, axis, labels and boxes.

The next function is named {\tt TionTrecSin2} and allow us to calculate the recombination times $t_{rec}$, defined as the time when the laser-ionized electron returns to the ion core, as a function of the ionization times $t_{ion}$, specified by the time when the bound atomic electron is put in the continuum due to the interaction with the laser electric field,  by solving numerically the Newton-Lorentz equation of motion for an electron moving in an oscillating electric field. We employ the Mathematica command {\tt NDSolve} in order to solve the second order differential equation with the adequate initial conditions and the package {\tt RootSearch} to find the times for which the electron returns to the parent ion, i.e. the recombination times. The variables needed for its calling are: 
\begin{itemize}
\item{The laser intensity, $I$, in W/cm$^{2}$}
\item{The laser wavelength, $\lambda$, in nm}
\item{The carrier envelope phase, $\phi$}
\item{The total number of laser cycles $n_p$}
\item{The number of points $n_{points}$ used in the discretization of the time variable. The time step is defined by $\delta t=n_p/n_{points}$}
\item{The range for ionization and recombination times (in optical cycles) to be used both for plots and data in the format $(t_{ion}^{min},t_{ion}^{max})$ and $(t_{rec}^{min},t_{rec}^{max})$} 
\item{Select the option to generate a Mathematica plot of $t_{rec}$ vs. $t_{ion}$ ({\tt Picture}), to export a pdf version of $t_{rec}$ vs. $t_{ion}$ plot ({\tt File}) or to save in a file a table with the numerical values of $t_{rec}$ vs. $t_{ion}$ ({\tt Data})}
\end{itemize}

For this case a typical example is as follows:

{\tt TionTrecSin2[1$\times$10$^{{\tt 14}}$, 1800, $\pi$, 10, 500, \{\{0, 3\},\{0, 3\}\},  \\
Option -> "File"]}

where we have exported a plot of $t_{rec}$ vs. $t_{ion}$ in the file {\tt 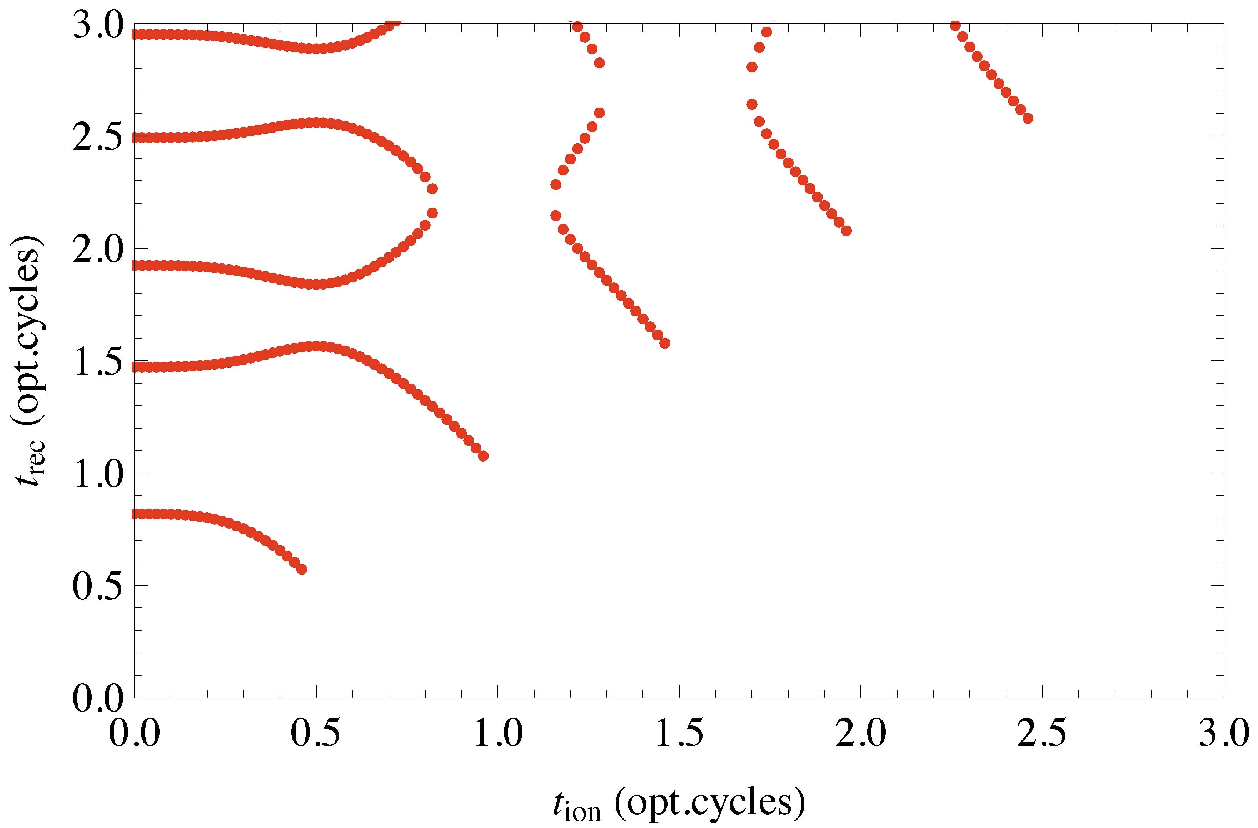}. This plot is presented in Fig.~1 where we have embedded directly the pdf file generated by the Mathematica notebook.

\begin{figure}[h]
\centering
\includegraphics[width=.75\textwidth]{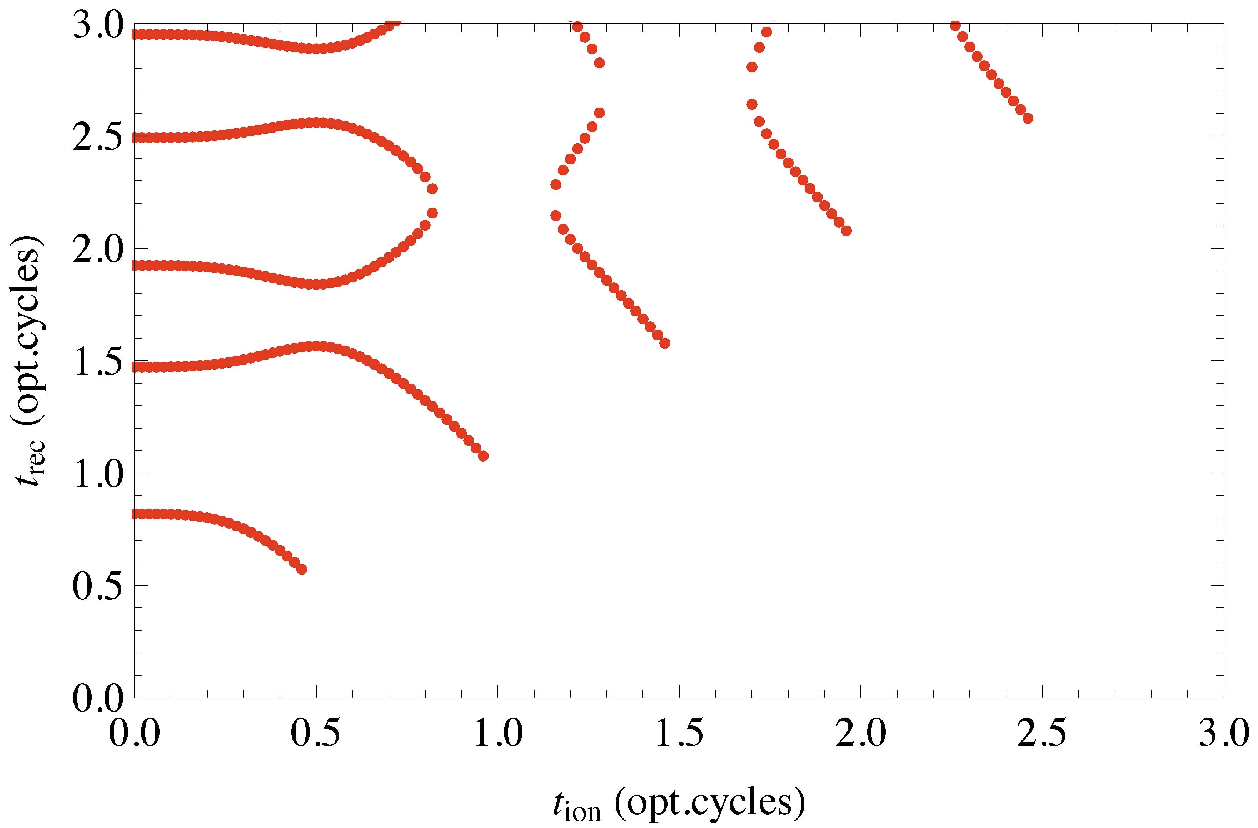}
\caption{Plot of a particular region, i.e. $0 < t_{ion} < 3$ and $0 < t_{rec} < 3$, of $t_{rec}$ vs. $t_{ion}$ for the
case of a sine-squared pulse (for the laser and additional parameters see the text).}
\label{fig1}
\end{figure}

Finally, the function named {\tt HHGEnergySin2} is the one defined to compute the the electron kinetic energy at the recombination time versus the ionization $t_{ion}$ and recombination $t_{rec}$ times. For this case the variables needed for its calling are: 
\begin{itemize}
\item{The laser intensity, $I$, in W/cm$^{2}$}
\item{The laser wavelength, $\lambda$, in nm}
\item{The carrier envelope phase, $\phi$}
\item{The total number of laser cycles $n_p$}
\item{The number of points $n_{points}$ used in the discretization of the time variable. The time step is defined by $\delta t=n_p/n_{points}$}
\item{The ionization potential $I_p$ of the atom or molecule under study (optional). The default value is $I_p=0.5$ a.u., which correspond to the ionization potential for the hydrogen atom.}
\item{The range for the ionization and recombination times (in optical cycles) ($x$-axis) and the values of the factor used for the energy range in terms of harmonic order or eV equivalent units ($y$-axis) to be employed for both plots and data in the format $(t_{ion,rec}^{min},t_{ion,rec}^{max})$ and $(f^{min},f^{max})$} 
\item{Select the option to generate a Mathematica plot of the kinetic energy at the recombination time versus the ionization $t{ion}$ and recombination $t_{rec}$ times ({\tt Picture}), to export a pdf version of the latter plot ({\tt File}) or to save in a file a table with the numerical values of the electron kinetic energy at the recombination time vs the ionization $t_{ion}$ and recombination $t_{rec}$ times ({\tt Data})}
\item{Select the units used for the electron kinetic energy at the recombination time. The options are {\tt Harm} for values of the kinetic energy in terms of harmonic order or {\tt eV} for values of the kinetic energy in terms of eV}
\end{itemize}

An example of a calling of this function yields

{\tt HHGEnergySin2[1$\times$10$^{{\tt 14}}$, 800, 0, 5, 500, \{\{0, 5\}, \{0, 1.5\}\}, \\
Option -> "File", Energy -> "Harm"]}

This function generates a file named {\tt 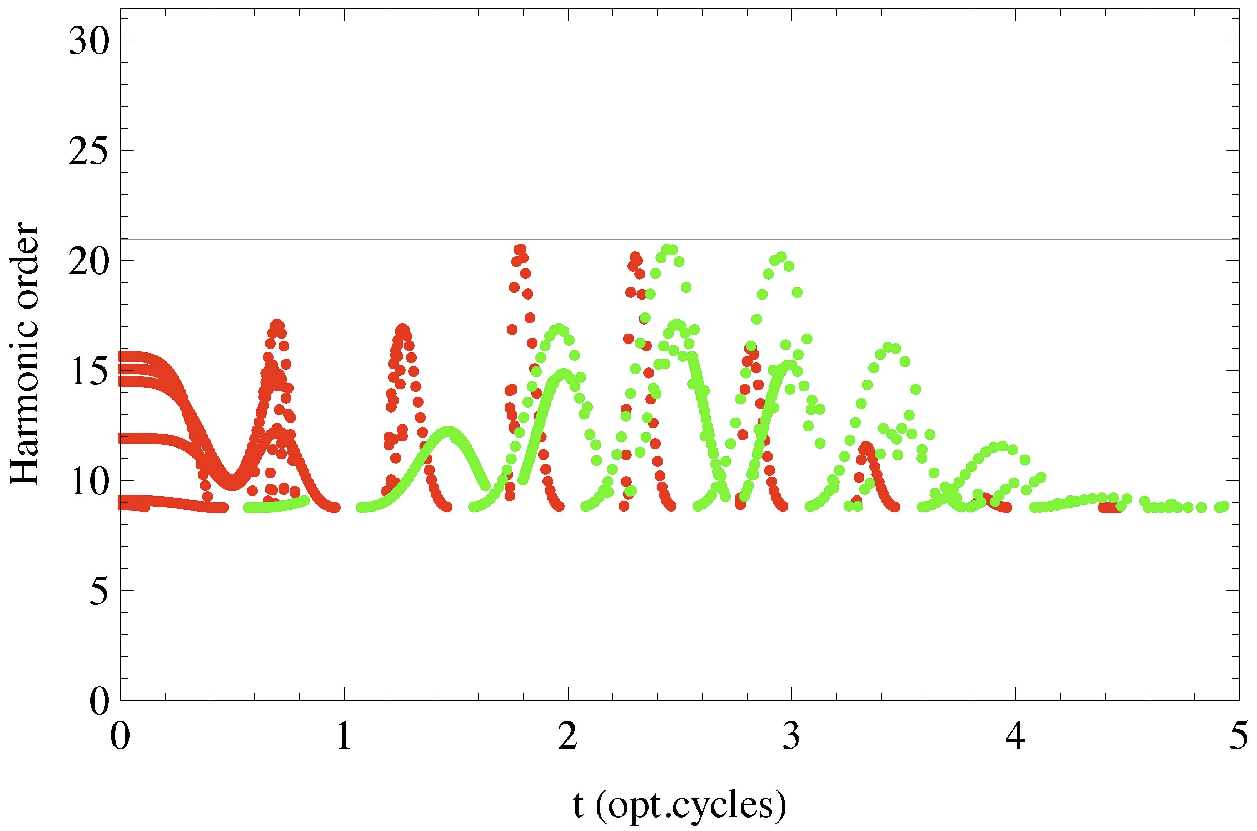} and it is shown in Fig.~2. Green (Red) circles correspond to the electron kinetic energy, in terms of harmonic order, at the recombination time as a function of the ionization (recombination) times. It includes a line indicating the value of the semiclassical cutoff~\cite{corkum,sfa} and we can observe our classical simulations exactly fulfill this limit. Note that, however, the ionization potential $I_p$ of the respective atom has been included in the electron kinetic energies calculations and for this reason all the HHG plots presented in the article do not start at 0 in the $y$-axis.

\begin{figure}[h]
\centering
\includegraphics[width=.75\textwidth]{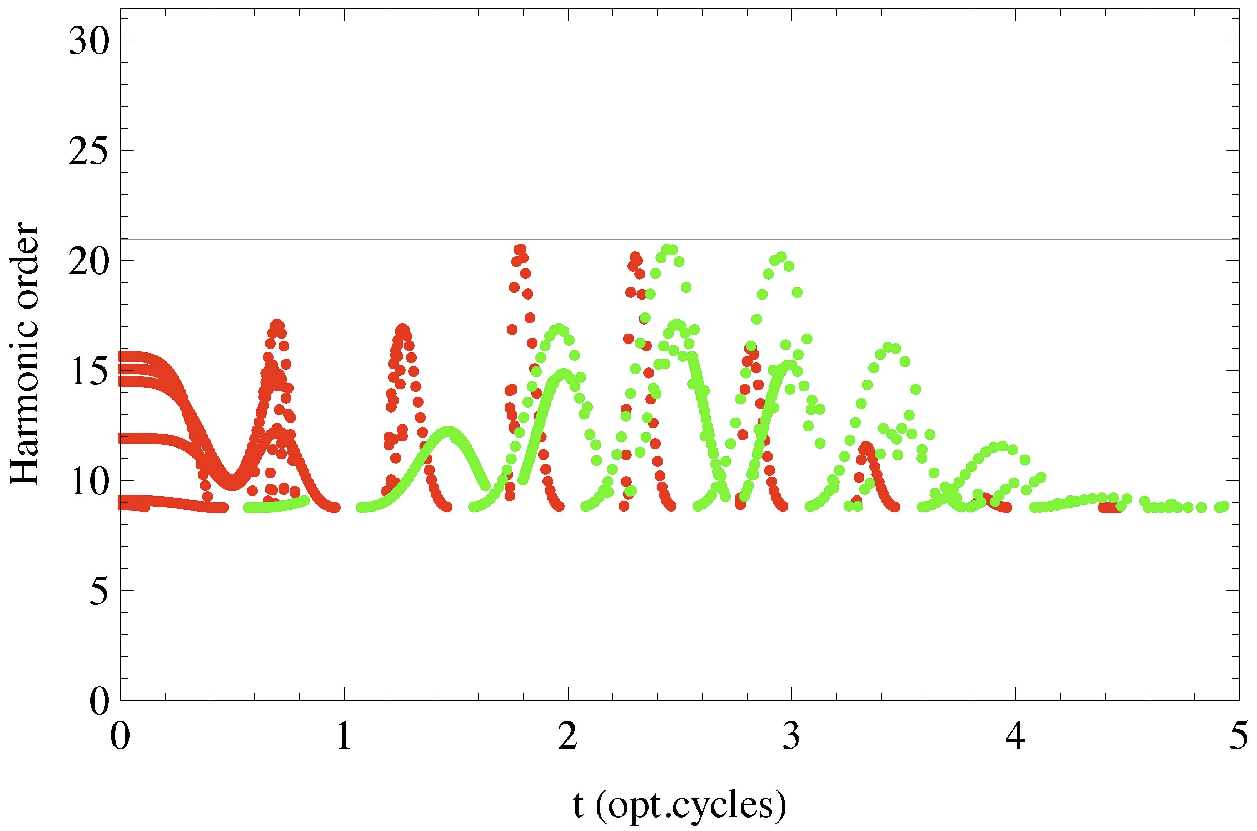}
\caption{Electron kinetic energy, in terms of harmonic order, at the recombination time as a function of the ionization time (green circles) and recombination time (red circles). The line at harmonic order 21 corresponds to the semiclassical cutoff (see the text for details). }
\label{fig2}
\end{figure}

In order to calculate HHG using gaussian laser pulses we only need to change the envelope of the laser electric field and to add up a couple of new parameters, namely the number of cycles full-width half maximum (FWHM) $n_{FWHM}$ and the total time duration of the simulation $n_s$. The rest of the functions are similar to the case of sine-squared pulses and they will not be described here (see the corresponding notebook for details). 

On the other hand, in order to compute HHG using trapezoidal laser pulses we only need to change the envelope of the laser electric field, as in the case of gaussian laser pulses, and to add up a couple of new parameters, namely the number of cycles of turn-on $n_{on}$, constant amplitude $n_p$ and turn-off $n_{off}$. The rest of the functions are similar to the case of sine-squared pulses and they will not be described here (see the corresponding notebook for details).

Finally, we have incorporated the possibility to calculate HHG using user defined arbitrary pulses. For instance, it is common to have experimentally available an array with the values of the laser electric field as a function of time and our approach allows to perform classical simulations with these particular kind of fields. The values of the laser electric field, in a.u., and time, in optical cycles, should to be saved in ascii format in a two columns file. In addition, an estimation of the laser wavelength needs to be provided (see the corresponding notebook for details).

\subsection{High-order harmonic generation (HHG) driven by spatial inhomogeneous fields}

In order to extend our contribution beyond the \textit{conventional} cases, we present in this sub-section studies of the HHG in atoms and molecules using spatially inhomogeneous cases. This kind of fields appears when a metal nanostructure, for instance nanotips or nanoparticles, is illuminated by a strong and short laser pulse. The main feature of these fields is that they present a spatial variation in the same scale where the electron dynamics takes place and consequently noticeable differences appear in the laser driven phenomena. Starting with the seminal experiment of Kim et al.~\cite{kimexp}, the activity on this field has been constant, lively and not without controversy (see e.g.~\cite{husakou,yavuz,ciappi1,tahir1,ciappiopt, ciappiati,ciappiann,fetic1,fetic2,sivis1,kimreply,ropersnew}). We point out, however, that the discussion about the experimental feasibility of this process falls out of the aim of the present work.

\subsubsection{The linear case}

We present here an example of HHG driven by a \textit{linear} spatially inhomogeneous field. For more details about the theoretical modeling with these fields and their implications see e.g.~\cite{ciappi1}. We employ a sine-squared laser pulse for this particular case, but it is straightforward to use another envelope to generate the theoretical results. The function named {\tt TionTrecLinear} allow us to calculate the recombination times $t_{rec}$ as a function of the ionization times $t_{ion}$ by solving numerically the Newton-Lorentz equation of motion for an electron moving in an oscillating and (linear) spatially varying electric field. A new variable, $\beta$, is needed in order to perform the classical simulations and it governs the inhomogeneity \textit{strength}~\cite{husakou,yavuz,ciappi1}. A typical example can be obtained by using:\\

{\tt TionTrecLinear[1$\times$10$^{{\tt 14}}$, 800, 0, 5, 500, 0.01, \{\{1, 2.5\}, \{1.5, 3\}\},  Option -> "File"]}

where we have set $\beta=0.01$ (note that $\beta$ will be the sixth parameter of the functions defined in this Section).

In Fig.~3 we present a plot of $t_{rec}$ versus $t_{ion}$ embedding directly the file {\tt 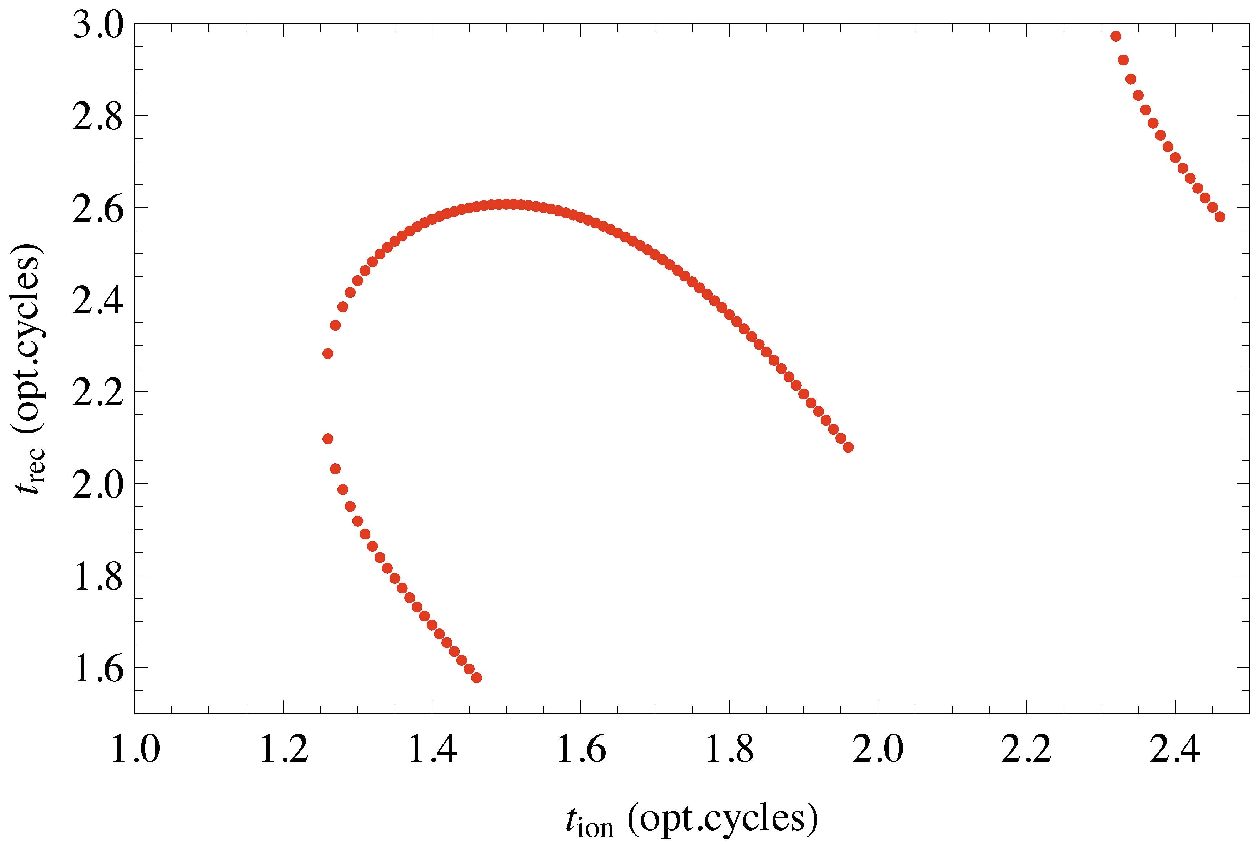} generated by the above defined function. The main difference with the homogeneous cases is the \textit{collapse} of the trajectories (for more details see~\cite{ciappi1}).

\begin{figure}[h]
\centering
\includegraphics[width=.75\textwidth]{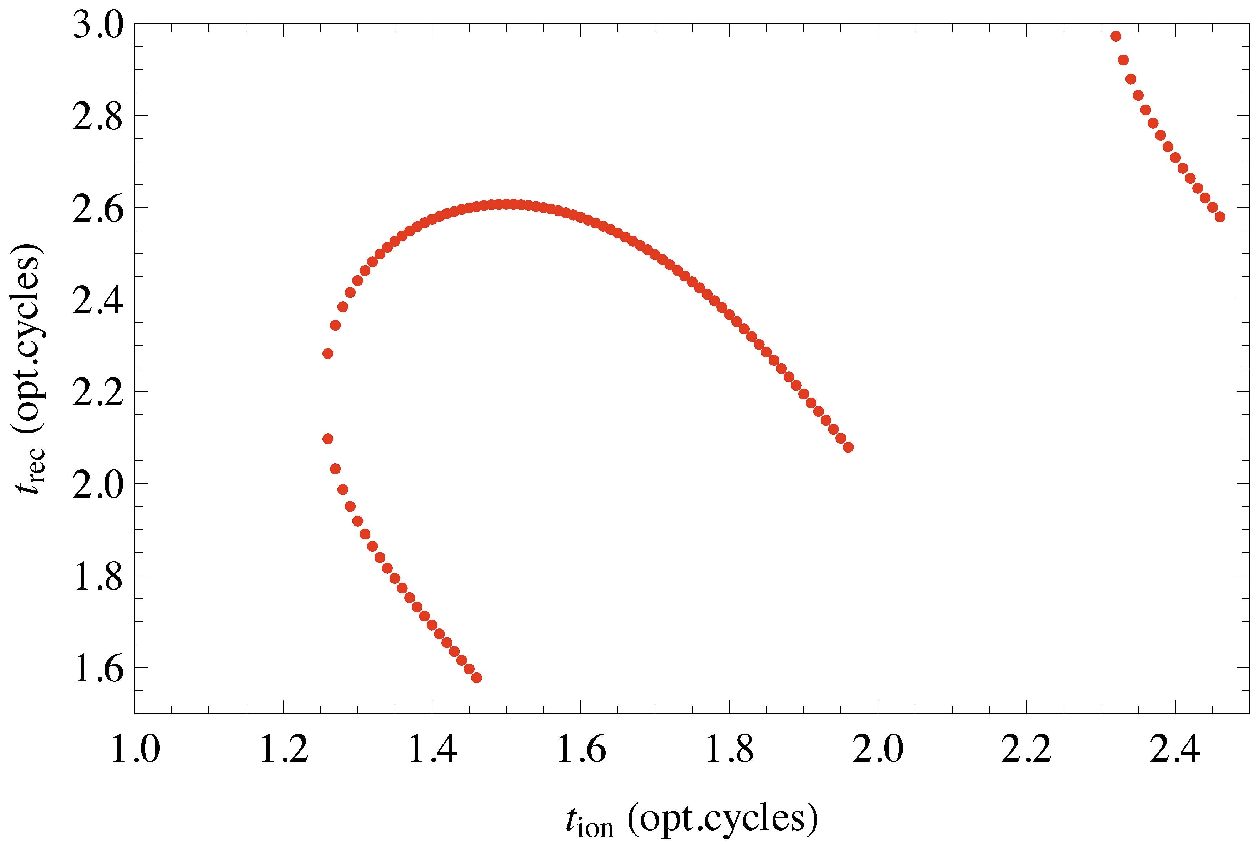}
\caption{Plot of a particular region, i.e. $1<t_{ion}<2.5$ and $1.5<t_{rec}<3$, of $t_{rec}$ vs. $t_{ion}$ for the case of a linear spatially nonhomogeneous field with $\beta=0.01$ (for the laser and atomic parameters see the corresponding notebook). }
\label{fig3}
\end{figure}

In addition, Fig.~4 shows the electron kinetic energy at the recombination time versus the ionization (green circles) and recombination times (red circles) in eV units, respectively. The plot is generated by using:

{\tt HHGEnergyLinear[1$\times$10$^{{\tt 14}}$, 800, 0, 5, 500, 0.01, \{\{0, 5\}, \{0, 3\}\}, \\
Option -> "File", Energy -> "eV"]}

where we employ $\beta=0.01$ and we have embedded directly the file named {\tt 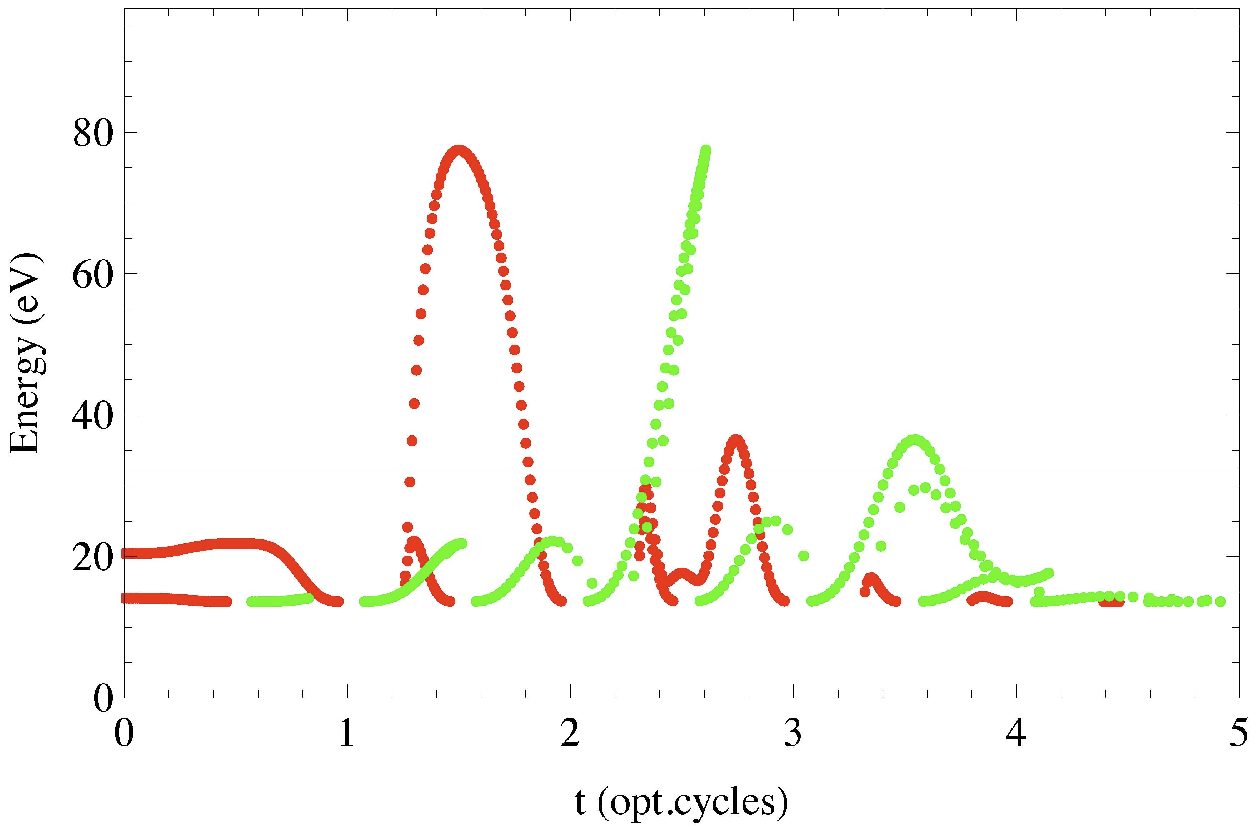}. As in the case of homogeneous fields, we include an horizontal line with the value of the semiclassical cutoff and we can observe that the HHG driven by this spatially (linear) nonhomogeneous fields reaches cutoff values of harmonic order far beyond this conventional limit (see e.g.~\cite{husakou,yavuz,ciappi1}).

\begin{figure}[h]
\centering
\includegraphics[width=.75\textwidth]{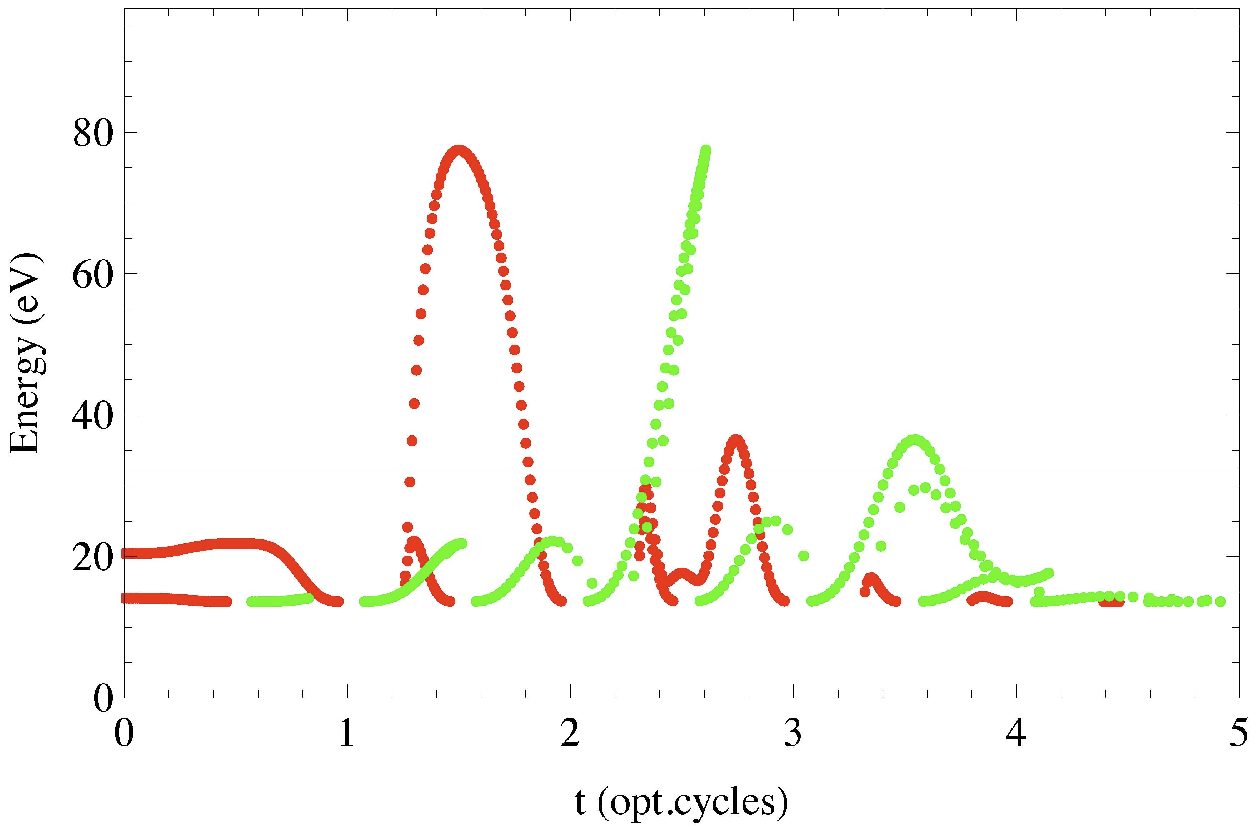}
\caption{Electron kinetic energy, in eV units, at the recombination time as a function of the ionization time (green circles) and recombination time (red circles)  for the case of a linear spatially nonhomogeneous field with $\beta=0.01$. The line at around 32.6 eV, value obtained from $21\times 1.55$, where 1.55 is the photon energy in eV for a laser wavelength $\lambda=800$ nm, corresponds to the semiclassical cutoff (see the corresponding notebook for details).}
\label{fig4}
\end{figure}

\subsubsection{The exponential case}

Another novel example of HHG driven by spatially nonhomogeneous field can be found in~\cite{tahirexp}. The plasmonic near-field generated when metal nanoparticles are illuminated by strong and short laser pulses has an exponential spatial variation and this represents a step forward compared with the simple linear case presented in the previous sub-section. In order to perform classical simulations of HHG driven by this particular spatially inhomogeneous fields we have developed functions including the parameter $\chi$ that governs the exponential inhomogeneity \textit{strength}. The actual value will be, for instance, function of the metal and the size of the nanoparticle employed (for details see~\cite{tahirexp}). The function named {\tt TionTrecExp} allow us to calculate the recombination times $t_{rec}$ as a function of the ionization times $t_{ion}$ by solving numerically the Newton-Lorentz equation of motion for an electron moving in an oscillating and (exponential) spatially varying electric field. A typical example, using the same laser parameters as in~\cite{tahirexp}, i.e. the laser intensity and wavelength are $I=2\times10^{13}$ W/cm$^2$ and $\lambda=720$ nm, respectively and the number of cycles of the sine-squared laser pulse is $n_p=5$, can be obtained with:\\

{\tt TionTrecExp[2$\times$10$^{{\tt 13}}$, 720, 0, 5, 500, 40, \{\{0, 1.5\}, \{0, 2\}\}, \\
Option -> "File"]}

where we have set $\chi=40$ (note that $\chi$ will be the sixth parameter of the functions defined in this Section). 

In Fig.~5 we present a plot of $t_{rec}$ versus $t_{ion}$ embedding directly the file {\tt 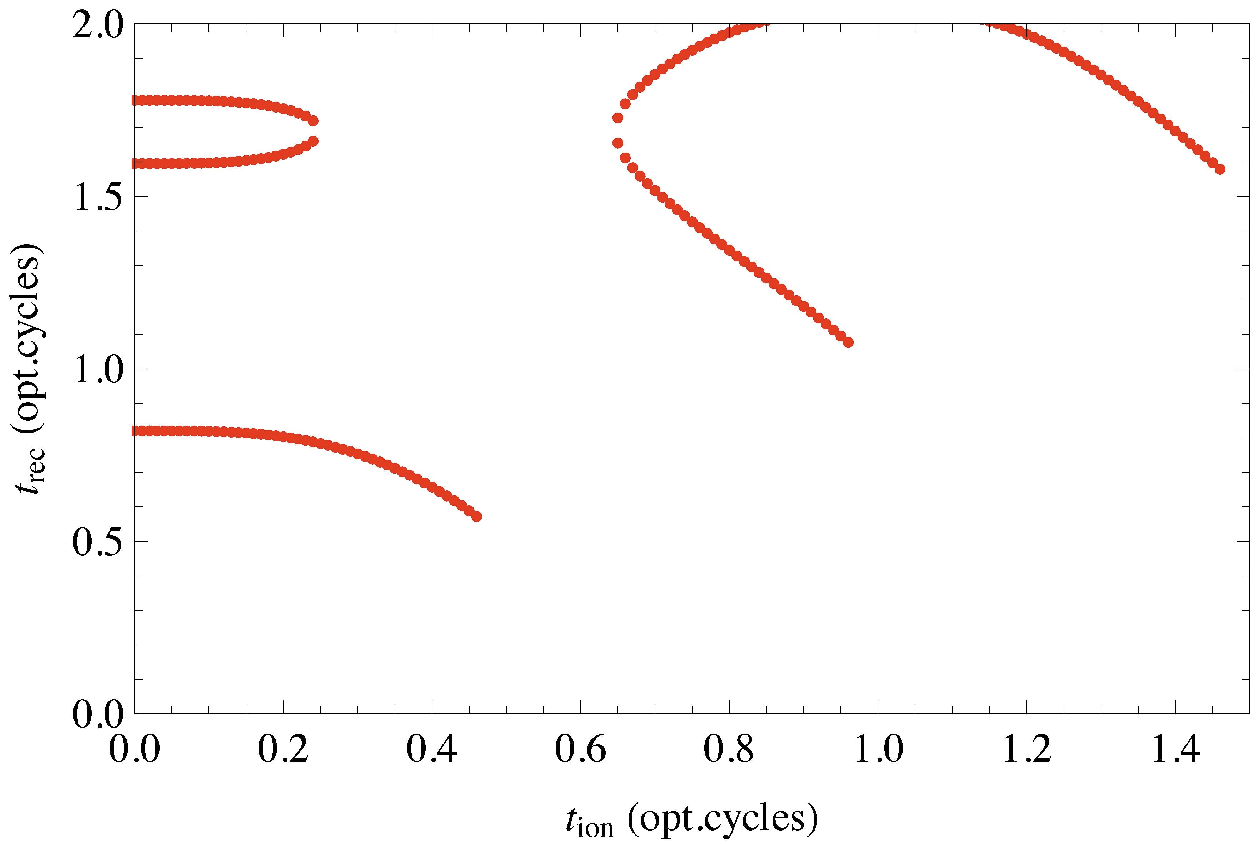} generated by the above defined function.

\begin{figure}[h]
\centering
\includegraphics[width=.75\textwidth]{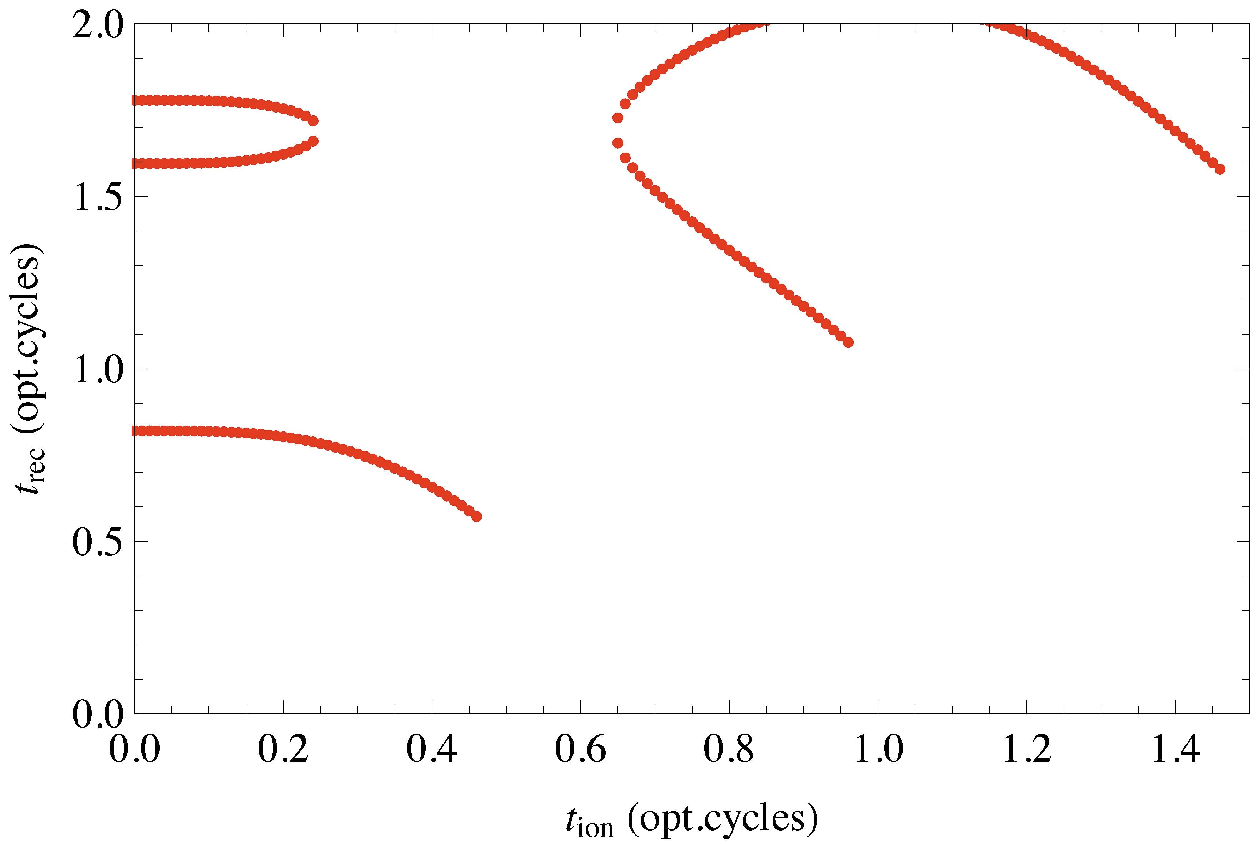}
\caption{Plot of a particular region, i.e. $0<t_{ion}<1.5$ and $0<t_{rec}<2$, of $t_{ion}$ vs. $t_{rec}$ for the case of an exponential spatially nonhomogeneous field with $\chi=40$ (for the laser parameters see the corresponding notebook). }
\label{fig5}
\end{figure}

Finally, Fig.~6 shows the electron kinetic energy at the recombination time versus the ionization (green circles) and recombination times (red circles) in harmonic order units. The function to generate this plot is as follows:

{\tt HHGEnergyExp[2$\times$10$^{{\tt 13}}$, 720, 0, 5, 1000, 40, \{\{0, 5\}, \{0, 4\}\}, \\
Option -> "File"]}

where we use $\chi=40$. The semiclassical cutoff with these parameters and for an H atom ($I_p=0.5$ a.u.) is $n_c\approx 9$ (or 15.5 eV) and is represented by the horizontal line in the figure. We can observe that the HHG driven by this particular (exponential) spatially nonhomogeneous field reaches cutoff values of harmonic order far beyond the conventional limit~\cite{tahirexp}.

\begin{figure}[h]
\centering
\includegraphics[width=.75\textwidth]{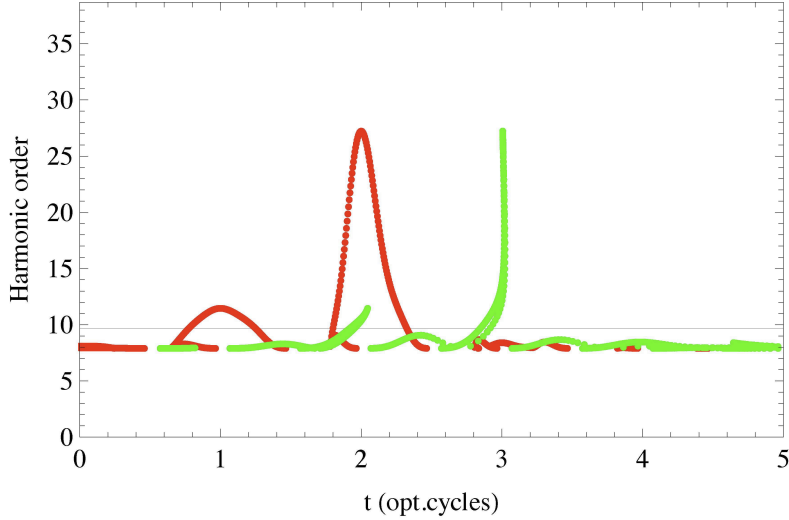}
\caption{Electron kinetic energy, in terms of harmonic order, at the recombination time as a function of the ionization time (green circles) and recombination time (red circles) for the case of an exponential spatially nonhomogeneous field with $\chi=40$. The line at harmonic order 9 corresponds to the semiclassical cutoff (see the corresponding notebook for details).}
\label{fig6}
\end{figure}
 
\subsection{High-order harmonic generation (HHG) driven by temporal synthesized fields}

In this sub-section we present HHG driven by few-cycle chirped pulses. We base our classical simulations in~\cite{carrera} and use similar
parameters in order to show the flexibility of our approach. The new functions include a new set of parameters in order to model a few-cycle chirped laser pulse. For instance, with the following function

 {\tt HHGEnergyChirp[1$\times$10$^{{\tt 14}}$, 800, 0, 5, 500, \{8.25, 200, 50\}, \\
 \{\{0, 5\}, \{0, 2\}\}, Option -> "Picture", Energy -> "Harm"]}

we compute the electron kinetic energy at the recombination time versus the ionization and recombination times in harmonic order, for a few-cycle ($n_p=5$) chirped laser pulse with a laser intensity $I=1\times10^{14}$ W/cm$^{2}$, $\lambda=800$ nm, $\beta=8.25$, $\tau=200$ a.u. and $t_0=0.25\tau$ a.u.

\subsection{High-order harmonic generation (HHG) driven by temporal and spatial synthesized fields}

Recently we have presented HHG in He atoms driven by temporal and spatial synthesized fields~\cite{joseprl} and we have shown it is possible to reach keV photons using laser intensities below the saturation limit. We do not enter into the details of this work and we only present here the main function and a representative graph.  An example using the function {\tt HHGEnergyTemporalSpatial} to compute the electron kinetic energy at the recombination time versus the ionization and recombination times is as follows:

{\tt HHGEnergyTemporalSpatial[1.4$\times$10$^{{\tt 15}}$, 800, 0, \{4, 6, 1.29\}, 0.002,\\
 500, 0.91,\{\{0, 6\}, \{0, 5\}\}, Option -> "File", Energy -> "eV"]}

and the generated plot is shown in Fig.~7~\cite{joseprl}.

\begin{figure}[h]
\centering
\includegraphics[width=.75\textwidth]{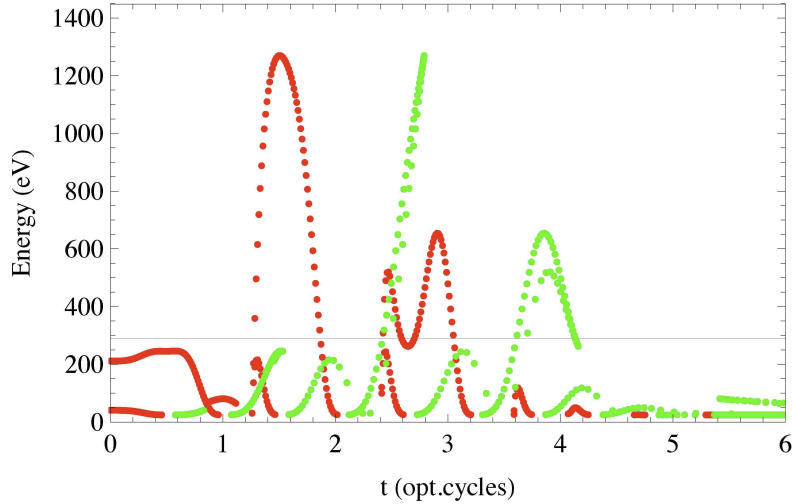}
\caption{Electron kinetic energy, in terms of eV at the recombination time as a function of the ionization time (green circles) and recombination time (red circles) for the temporal and spatial synthesized laser pulse studied in~\cite{joseprl}. The line at around 290 eV, value obtained from $187\times 1.55$, where 1.55 is the photon energy in eV for $\lambda=800$ nm, corresponds to the semiclassical cutoff (see the corresponding notebook for details).}
\label{fig7}
\end{figure}

\section{Above-threshold ionization (ATI)}

\subsection{Above-threshold ionization (ATI) driven by spatial homogeneous fields}
As in the case of HHG we start by modeling above-threshold ionization (ATI) in atoms and molecules with sine-squared laser pulses. We avoid to repeat the detailed description of the latter case and also for the gaussian and trapezoidal laser pulses ones (see the corresponding HHG cases and the notebooks for further details), and we only show a prototypical example. We will include, however, ATI driven by spatial and temporal synthesized fields, in order to show how this new kind of fields modifies substantially the energy limits and the electron trajectories (for more details about ATI driven by spatial inhomogeneous fields see e.g.~\cite{ciappiati,tahirati,ciappiati2}).

\subsubsection{Above threshold ionization (ATI) with sine-squared laser pulses}
Similarly to the case of HHG we need a couple of initialization commands before the definition of the Mathematica functions, namely, \\
{\tt  SetDirectory[NotebookDirectory[]]} \\
and \\
{\tt <<RootSearch.m}\\
The first one sets the working directory to the {\tt  NotebookDirectory[]}, i.e. where the current notebook is located. The second initialization line loads the package {\tt  RootSearch.m} using the {\tt  Get} ($<<$) command. Please note that the {\tt  RootSearch.m} package should to be located in the same directory where the notebooks are. For details about the {\tt  RootSearch.m} see e.g.~\cite{rootsearch}.

The functions {\tt PulseSin2} and {\tt TionTrecSin2} are identical to the ones defined for the case of HHG and we include them in the notebook only for completeness. The main function to compute the electron kinetic energies for both the direct and rescattered electrons is {\tt ATIEnergySin2} and a typical example is as follows:

{\tt ATIEnergySin2[1$\times$10$^{{\tt 14}}$, 800, 0, 5, 3000, \{\{0, 5\}, \{0, 12\}\}, \\
Option -> "File", Energy -> "Up"]}

This function generates a file named {\tt 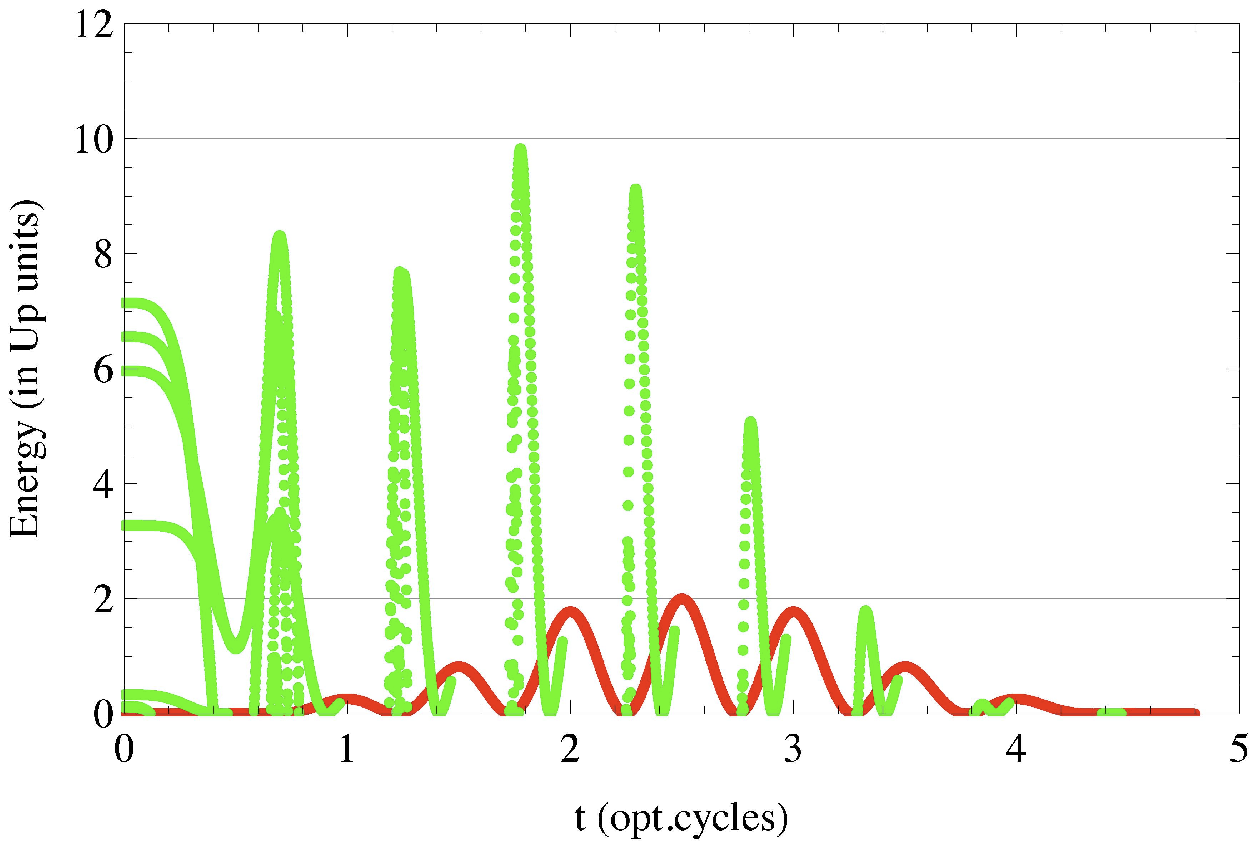} and is shown in Fig.~8. Red (Green) circles correspond to the electron kinetic energy, in terms of $U_p$ units ($U_p=I/4\omega^2$ is the ponderomotive energy with $I$ the laser intensity and $\omega$ its frequency), of the direct (rescattered) electron as a function of the ionization time. It includes lines indicating the values of the classical electron kinetic energies limits for both the direct, $2U_p$, and rescattered, $10 U_p$, electron~\cite{miloreview}. We can observe our classical simulations exactly fulfill these limits.

\begin{figure}[h]
\centering
\includegraphics[width=.75\textwidth]{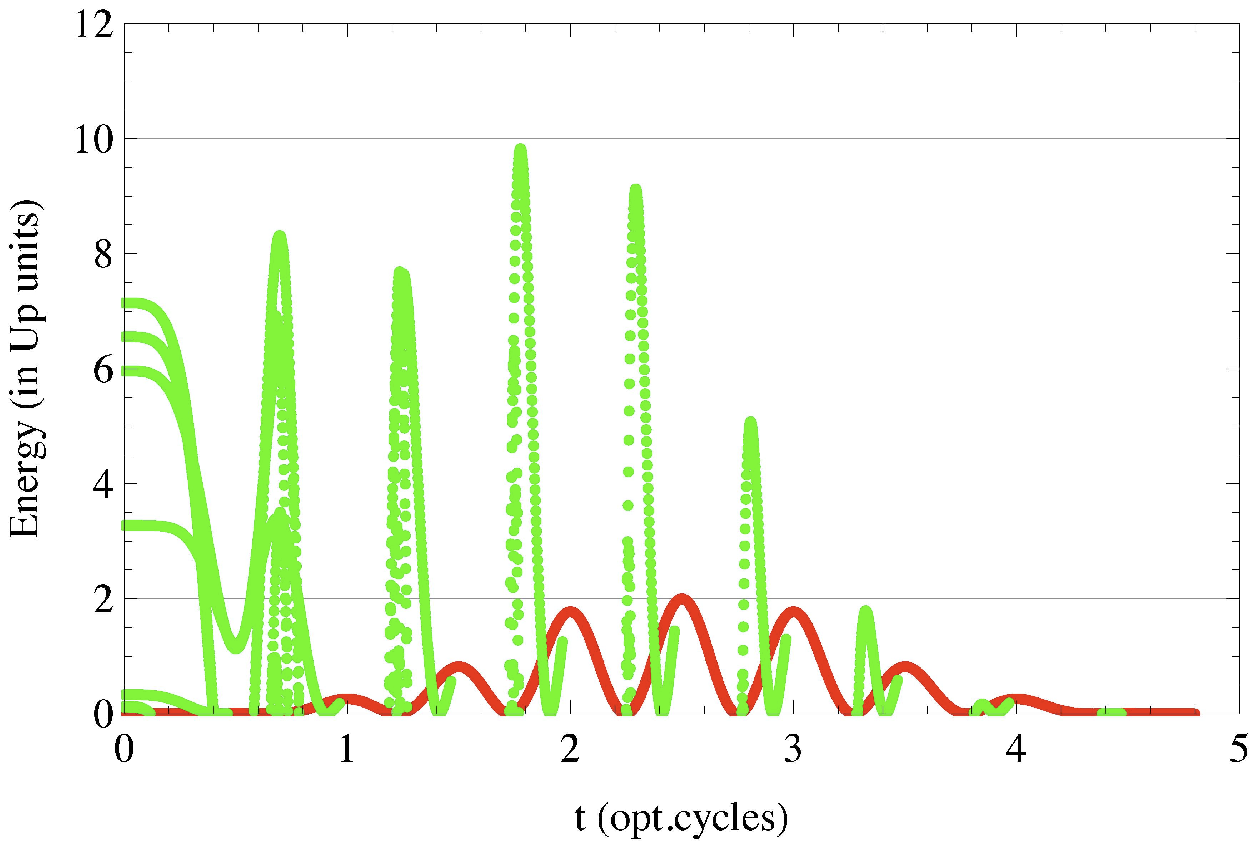}
\caption{Electron kinetic energies for the direct (red circles) and rescattered (green circles) electron, in units of $U_p$, as a function of the ionization time. The lines at $2$ and $10$ are the maximum electron kinetic energies for the direct and rescattered electron, respectively (see the text and~\cite{miloreview} for details). }
\label{fig8}
\end{figure}

As in the case of HHG, to calculate ATI using gaussian laser pulses we only need to change the envelope of the laser electric field and to add up a couple of new parameters, namely the number of cycles full-width half maximum (FWHM) $n_{FWHM}$ and the total time duration of the simulation $n_s$. The rest of the functions are similar to the case of sine-squared pulses and they will not be described here (see the corresponding notebook for details). 

On the other hand, in order to compute ATI using trapezoidal laser pulses we only need to change the envelope of the laser electric field, as in the case of gaussian laser pulses, and to add up a couple of new parameters, namely the number of cycles of turn-on $n_{on}$, constant amplitude $n_p$ and turn-off $n_{off}$. The rest of the functions are similar to the case of sine-squared pulses and they will not be described here (see the corresponding notebook for details).

Finally, we have incorporated the possibility to calculate ATI using user defined arbitrary pulses. For instance, it is common to have experimentally available an array with the values of the laser electric field as a function of time and our approach allows to perform classical simulations with these particular kind of fields. The values of the laser electric field, in a.u., and time, in optical cycles, should to be saved in ascii format in a two columns file. In addition, an estimation of the laser wavelength needs to be provided (see the corresponding notebook for details).

\subsection{Above-threshold ionization (ATI) driven by spatial inhomogeneous fields}

We present in this sub-section studies of the ATI in atoms and molecules using spatially inhomogeneous cases. As was pointed out in Section 2.2 this kind of fields appears when a metal nanostructure, for instance nanotips or nanoparticles, is illuminated by a strong and short laser pulse. Similarly to the HHG case, the ATI features will be strongly modified (for more details see~\cite{ciappiati,tahirati,ciappiati2}).  

\subsubsection{The linear case}

This sub-section is the counterpart of Section 2.2.1 for the ATI phenomenon and consequently we do not give more details about the Mathematica functions We only want to show how the (linear) spatial variation of the laser electric field, modifies substantially the electron kinetic energy of both the direct and rescattered electron, as can be seen in Fig.~9.

\begin{figure}[h]
\centering
\includegraphics[width=.75\textwidth]{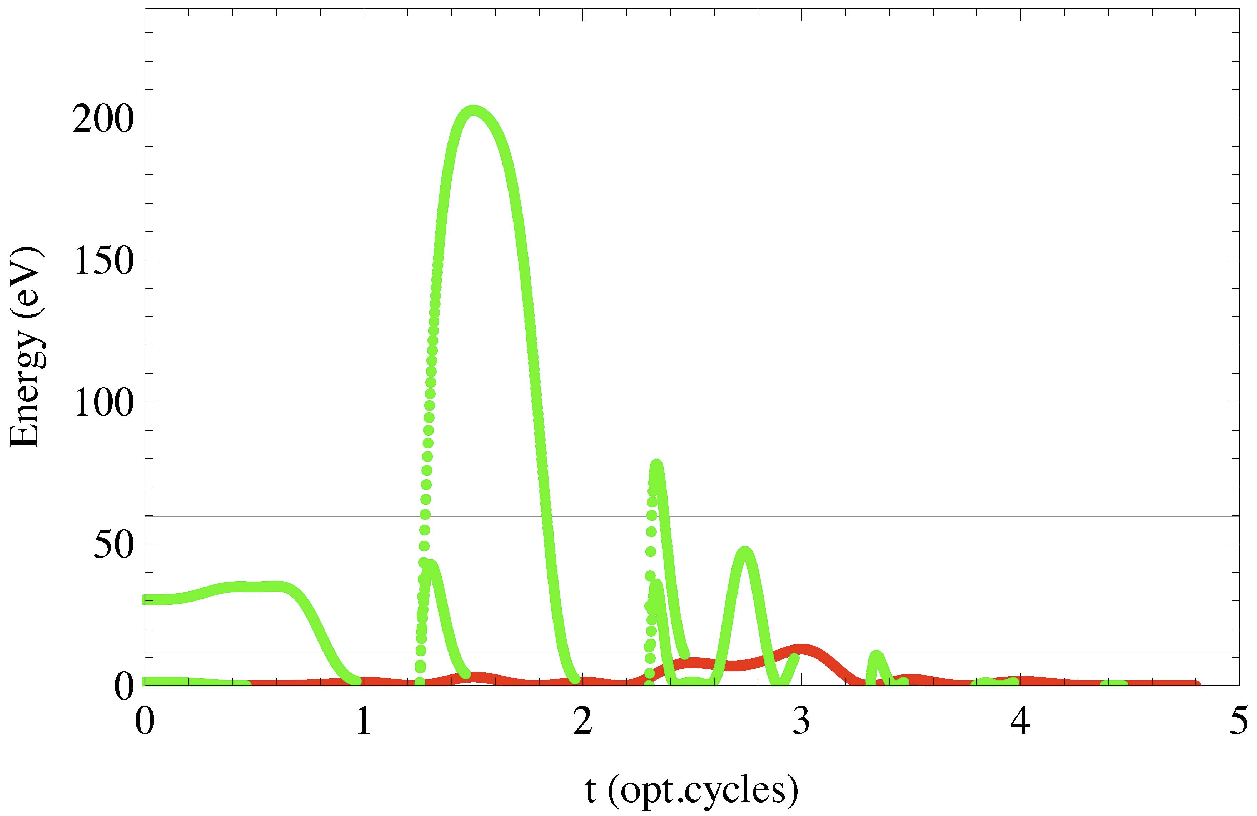}
\caption{Electron kinetic energies for the direct (green circles) and rescattered (red circles) electron, in eV, as a function of the ionization time. The laser parameters are $I=1\times10^{14}$ W/cm$^{2}$ and $\lambda=800$ nm. We use a $\beta=0.01$ for the inhomogeneity strength. The lines at $\approx 12$ eV and $\approx 60$ eV are the maximum electron kinetic energies for the direct and rescattered electron, respectively (see the text and~\cite{ciappiati,tahirati,ciappiati2} for details). }
\label{fig9}
\end{figure}
%

\subsubsection{The exponential case}

This sub-section is the counterpart of Section 2.2.2 for the ATI phenomenon and consequently we do not give more details about the Mathematica functions for this particular case. As in the previous section we only want to show how the (exponential) spatial variation of the laser electric field, modifies substantially the electron kinetic energy of both the direct and rescattered electron, as can be seen in Fig.~10. The most important difference with the previous cases can be observed in the electron kinetic energy for the direct electron $E_d$, that now can reach values comparable to the kinetic energy for the rescattered electron $E_r$.  
%
%
\begin{figure}[h]
\centering
\includegraphics[width=.75\textwidth]{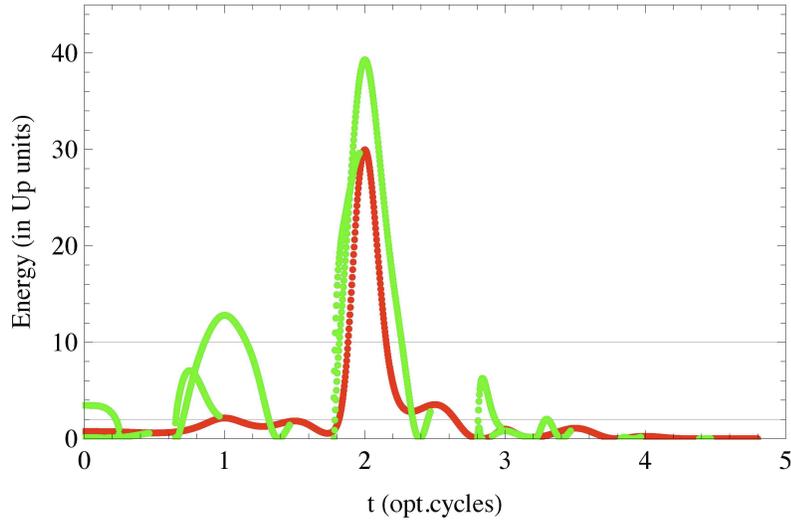}
\caption{Electron kinetic energies for the direct (red circles) and rescattered (green circles) electron, in units of $U_p$, as a function of the ionization time. The laser parameters are $I=2\times10^{13}$ W/cm$^{2}$ and $\lambda=720$ nm. We use a $\chi=40$ for the inhomogeneity strength (exponential). The lines at $2$ and $10$ are the maximum electron kinetic energies for the direct and rescattered electron, respectively (see the text and the corresponding notebook for details). }
\label{fig10}
\end{figure}

\subsection{Above-threshold ionization (ATI) driven by temporal synthesized fields}

This sub-section is the counterpart of Section 2.3 for the ATI phenomenon and consequently we do not give more details about the Mathematica functions. As in the previous sections we only want to show how the temporal synthesized fields modifies substantially the electron kinetic energy of both the direct and rescattered electron, as can be seen in Fig.~11. The ATI phenomenon driven by few-cycle chirped pulses was studied in detail in~\cite{atichirp}.

\begin{figure}[h]
\centering
\includegraphics[width=.75\textwidth]{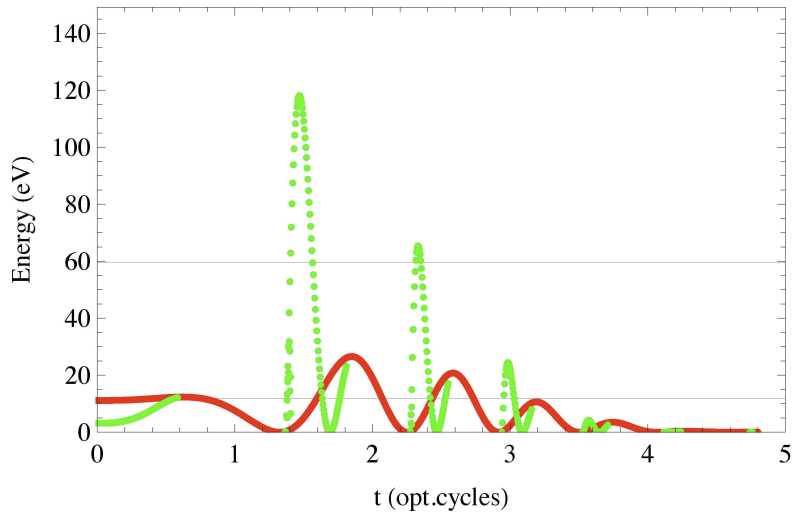}
\caption{Electron kinetic energies for the direct (red circles) and rescattered (green circles) electron, in eV, as a function of the ionization time. The laser parameters are $I=1\times10^{14}$ W/cm$^{2}$ and $\lambda=800$ nm. We use the few-cycle chirped laser pulse defined in Section 2.3. The lines at $\approx 12$ eV and $\approx 60$ eV are the maximum electron kinetic energies for the direct and rescattered electron, respectively (see the text and the corresponding notebook for details). }
\label{fig11}
\end{figure}
 

\subsection{Above-threshold ionization (ATI) driven by temporal and spatial synthesized fields}

This sub-section is the counterpart of Section 2.4 for the ATI phenomenon and consequently we do not give more details about the Mathematica functions. As in the previous sections we only want to show how a temporal and spatial synthesized field modifies substantially the electron kinetic energy of both the direct and rescattered electron, as can be seen in Fig.~12. We can observe from these plots it is possible to reach electron kinetic energies in the keV regime.

\begin{figure}[h]
\centering
\includegraphics[width=.75\textwidth]{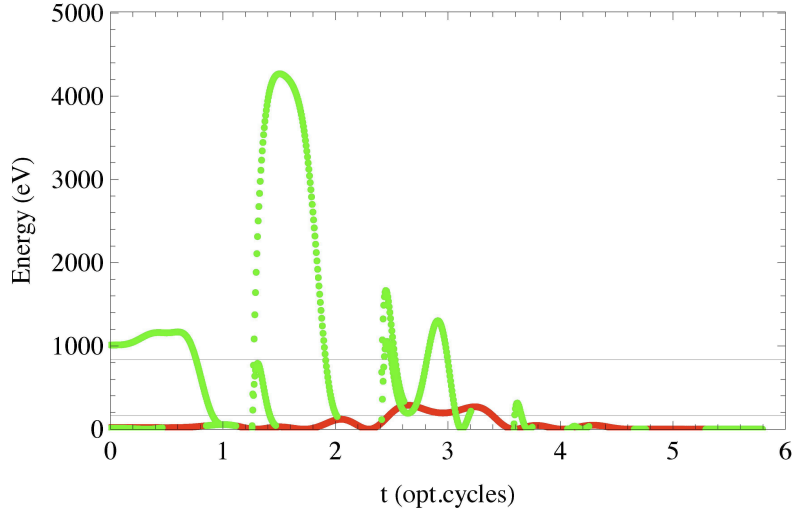}
\caption{Electron kinetic energies for the direct (red circles) and rescattered (green circles) electron, in eV, as a function of the ionization time. The laser parameters are $I=1.4\times10^{15}$ W/cm$^{2}$ and $\lambda=800$ nm (see~\cite{joseprl} for details). We use the spatial and temporal synthesized laser pulse defined in Section 2.4. The lines at $\approx 170$ eV and $\approx 850$ eV are the maximum electron kinetic energies for the direct and rescattered electron, respectively (see the text and the corresponding notebook for details). }
\label{fig12}
\end{figure}
 
 
\section{Conclusions and perspectives}

A set of Mathematica notebooks for the classical simulation of strong field physics phenomena has been presented. 
In particular we have studied the high-order harmonic generation (HHG) and above-threshold ionization (ATI). By predicting with great precision the cutoff positions both in the HHG and ATI processes, the classical approach constitutes a fast and useful tool as it was shown throughout  the article. A variety of laser pulses envelopes are used in order to model the high-order harmonic and above-threshold ionization spectra features. The notebooks are based on functions which allow to export both pre-formatted graphs and raw data that could be used to generate plots and analyzed using different graphical tools. We have included an ample set of laser electric fields, namely electric fields that are spatially inhomogeneous. This kind of field appear when metal nanosystems are illuminated by short and intense laser pulses and have the property to spatially vary in the same scale where the electron dynamics takes place. Additionally, we have incorporated in the simulations the possibility of using spatial synthesized fields. Finally, we have combined both the temporal and spatial synthesized fields, in order to complete the theoretical picture and to allow to understand the underlying physics of these novel laser driven sources.

\ack

We acknowledge the financial support of the MICINN
projects (FIS2008-00784 TOQATA, FIS2008-06368-C02-01,
and FIS2010-12834), ERC Advanced Grant QUAGATUA, EU IP SIQS, the
Alexander von Humboldt Foundation, and the Hamburg Theory
Prize (M.L.). This research has been partially supported
by Fundaci\'o Privada Cellex.
J. A. P.-H. acknowledges support from Spanish
MINECO through the Consolider Program SAUUL
(CSD2007-00013) and research project FIS2009-09522,
from Junta de Castilla y Le\'on through the Program for
Groups of Excellence (GR27) and from the ERC Seventh
Framework Programme (LASERLAB-EUROPE, Grant
No. 228334). We thank Samuel Marskon for useful comments and remarks.




\end{document}